The Drake Equation at 60: Reconsidered and Abandoned

John Gertz[1][2]

**Abstract**: Each of the individual factors of the Drake Equation is considered. Each in turn is either abandoned or redefined and finally reduced to a single new factor, $f_d$, the fraction of technological life that is detectable by any means. However, neither the Drake Equation, nor its replacement, can actually solve for N. Only a vibrant SETI program and, ultimately, contact with an alien civilization might result in the determination of N.

**Keywords**: SETI, ET, Drake Equation, Habitable Zone

## 1.    INTRODUCTION

On April 8, 1960, Frank Drake ushered in the era of modern SETI research when he pointed the Green Bank Telescope at two nearby Sun-like stars, Tau Ceti and Epsilon Eridani [1]. A year later, Drake, along with Carl Sagan, organized a three-day conference on SETI, the first such conference ever held on the topic. It convened on November 1, 1961 at the Green Bank Observatory in West Virginia. In an effort to organize the agenda for the meeting, Drake jotted down a heuristic meant to guide the discussions. Drake's doodle on the back of the proverbial envelope grew in stature to become the almost sacred "Drake Equation":

$N = R^* f_p n_e f_l f_i f_c L$             (1)

Where,

$R^*$ = the rate of Sun-like star formation in the Milky Way.

$f_p$ = the fraction of stars harboring planets.

$n_e$ = the average number of those planets per star system that are Earth-like, in the sense that they are capable of harboring life.

$f_l$ = the fraction of those planets that actually do bear life of any sort.

[1] *Zorro Productions, Berkeley, California*
[2] Correspondence address: Zorro Productions, 2249 Fifth Street, Berkeley, CA, 94710. Tel: (510) 548-8700. *Email address*: jgertz@zorro.com

$f_i$ = the fraction of life bearing planets among whose life forms exists one or more of at least human-like intelligence such that it is capable of transmitting radio, infrared, or optical signals.

$f_c$ = the fraction of those intelligent species who are attempting to communicate with Earth at our present time.

L = the length of time these civilizations persist in transmitting to Earth. Alternatively, and more broadly, L is taken to be the average life expectancy of ET civilizations.

The product of these factors yields N, the number of alien civilizations that are technologically competent to transmit electromagnetic signals and are actually transmitting to Earth at this time.

The Drake Equation has been called, probably correctly, the second most famous equation in science, overshadowed only by Einstein's $E=Mc^2$. Drake, himself downplayed the importance of the equation, noting that "it didn't take any deep intellectual effort or insight on my part" [2]. While an admirable attempt to parse the problem as conceived at the outset of modern SETI, much has been learned since then, while much remains as unknowable today as it was in 1961. After 60 years, a factor by factor reconsideration may be appropriate.

## 2. R* ➔ $n_s$

Drake immediately got off on the wrong foot by starting his equation with a virtually useless parameter, R*, the number of Sun-like stars born each year in the Milky Way. Its value is fairly well understood to be approximately one star per year at this time in the galaxy's evolution. However, knowing this number does not help us at all to understand the current distribution of communicating ET civilizations.

### 2.1 Limitations of Drake's R*

2.1.1 The rate of star formation changes over time. It is much lower today than it was in the past, having peaked about 10 billion years ago.

2.1.2 Drake limited his definition to Sun-like stars. However, there is good reason to believe that ET might be found in star systems that are decidedly un-Sun-like. The birthrate of all stars is much higher than for Sun-like stars and is currently closer to 10 than to 1.

2.1.3 R* assumes that the signals that we might detect will derive from within our own galaxy. However, transmissions may in fact be intergalactic. A very advanced civilization might be able to harness so much energy that they are able to construct transmitters that might illuminate an entire remote galaxy at once with a detectable flux.  Thus, instead of transmitting to one star at a time with the hope that that one star might both harbor a sister civilization and that that civilization might be looking exactly in its direction, it can send the same message all at once to, in the case of the Milky Way, some 200-400 billion stars.

2.1.4   The number of civilizations in the Galaxy, N, is not related to how many stars formed in the past year, or even past billion years.  Thus, R* should not appear in the Drake Equation.  What does matter is the total number of current stars that can serve as hosts for life.

## 2.2   Replacing R* with Number of Stars ($n_s$)

R* might be replaced by $n_s$, denoting the number of candidate stars in the Milky Way. We should be able to calculate this number within a factor of 2 or 3 if we can agree on which stars to include:

2.2.1   Surely, we can exclude A, B, and O stars, as these live very short lives, ending long before they could possibly develop their own intelligent life forms.  Taken together, they represent less than 1% of stars, so whether to include or exclude them makes no practical difference in the calculation of N.

2.2.2   The first technological species evolved on Earth only 4.55 billion years after its birth.  We have no idea whether this is typical.  We can only say that any star system that is older than 4.55 billion years should be a candidate target for SETI.  Because star formation peaked 8 – 11 billion years ago, approximately 96% of all stars are older than Sol.  To conduct a SETI search in the most efficient manner possible one might avoid the very youngest stars.  However, stars that are younger than Sol cannot be confidently ignored because civilizations might take less than 4.5 billion years to emerge and even very young planetary systems might be populated by aliens who have rushed to colonize their virgin turf.

2.2.3   White dwarfs, representing about 10% of all stars, might be subtracted from the total.  However, an alien civilization residing in orbit around a dying star has only three choices: (a) it can passively submit to extinction; (b) it can move away from its home star; or (c) it can find a way to persist where it is.  If the best option is to stay where it is, it will be forced to persist in orbit around a white dwarf, and therefore white dwarfs are viable SETI targets [3].

2.2.4   Drake was interested in the rate of G-dwarf births because he intuited that ours is the best of all possible worlds, and that therefore ET must also reside around a G-dwarf. Early SETI theorists broadened their target lists to include the somewhat hotter F-dwarfs and somewhat cooler K-dwarfs, but no consideration at all was given to the most numerous class of stars, cool red M-dwarfs. Since they represent 2/3 to 3/4 of the stars in our neighborhood of the Milky Way, eliminating them from consideration would significantly reduce $n_s$. It was reasoned that life could not exist on planets in orbit around M-dwarfs because their habitable zones (HZ) are too thin, and the chances of a rocky planet found within their HZs is therefore small. However, Kepler data indicates that there may actually be more rocky planets within the HZs of M-dwarfs than G-dwarfs [4]. Indeed, rocky planets have been discovered within the HZs of M-dwarfs, including in orbit around Earth's closest neighbor, Proxima Centauri, as well around TRAPPIST-1, an M-dwarf at 39 light years with 7 known rocky planets, 3 of which are located within its HZ. M-dwarfs emit X-ray and UV flares that might sterilize a planet and strip it of its atmosphere. Alternatively, their atmospheres might prevent the worst damage from radiation and/or life forms might depend on flares to drive mutations necessary for evolution. We simply do not know.

2.2.5   The enumeration of stars ignores the possibility that ET may be post-biological or AI beings, not constrained to residing on planets.   I have elsewhere considered the possibility that the entire Milky Way may be wired for communication by means of alien AI probes located around many, most or all stars [5,6,7,8].   If such be the case, then ET's presence would be ubiquitous.  There are caveats.  Even if ET probes or and communications nodes are to be found in orbit around every star system, this tells us nothing about the true nature of Drake's N, the number of extant civilizations.  Just a handful of civilizations might be represented on the galactic Internet.  Alternatively, that Internet might contain a vast library of information regarding civilizations that are themselves long extinct.  The informational content of the galactic Internet might only grow over deep time, regardless of whether the number of contributing civilizations increases or diminishes.  Taken to the extreme, all other civilizations might now be extinct while the self-replicating and self-repairing nodal system they bequeathed endures.  The second caveat is that the whole point of a nodal system is to increase the efficiency of communication by relying on short range spokes.  If nodes are located around most stars, then we might hear from or intercept a communication only from a very nearby star.  If that is the case (leaving aside for the moment the possibility of a detection of a techosignature not associated with an intentional transmission to Earth), $n_s$ is a very small number, including only those stars closest to Sol.  Paradoxically, if there is only one probe or node that we might hear from, then N = 1, but nonetheless, the chances of making contact with that N is unity, as soon as it deigns to break its silence and transmit to Earth.   Moreover, that probe or node would represent a gateway into the entire galactic knowledge base representing input from a vastly larger set of civilizations, be they currently extant or extinct.

**2.3     Replacing R* with Number of Spots on the Sky (n$_s$)**

However, a complete list of viable SETI targets should include non-stellar sources (since "star" and "source" both begin with "s", we can keep the n$_s$ nomenclature), among which are:

- We might detect the waste heat generated by or an intentional EM transmission from a K-III civilization in another galaxy.

- ET might reside on rogue or unbound planets in interstellar space that have been flung out of their solar systems of origin and are now free floating in interstellar space.  When a solar system is formed its planets coalesce into random orbits, many of which are not gravitationally stable.  Bigger, Jupiter-sized planets, tend to fling the smaller, Earth-sized planets out of the system, which are usually the losers in the gravitational tug of war.   It is currently estimated that there are hundreds of billions, and perhaps even a trillion of these rogue planets within the Milky Way.  If such a rogue planet possesses an internal heat sources, such as seafloor geothermal vents, they could have spawned life even in the absence of stellar flux.  Alien civilizations might have intentionally colonized rogue planets, perhaps as they approached their own solar systems, in an attempt to ward off the devastation that would be wrought by their own star's inevitable demise.  Alternatively, an alien civilization that values computation above all else, might value their ultracold, computationally efficient temperatures.

- An alien world ship cruising the depths of interstellar space might be detected.

- An alien probe within our own solar system might be detected.

Since a non-stellar detection might be made at any set of coordinates, every single point in space constitutes a viable target.  Therefore, it would seem that there are an infinite number of spots on the sky to observe.  However, the introduction of an infinity into the Drake Equation would be disastrous, leading immediately to the solution:

$$N = \infty \qquad (2)$$

This is absolutely impossible, since the visible universe, while being very big, is not infinite and cannot contain an infinite number of ET civilizations.  The conundrum is in part resolved by recognizing that no telescope observes only an infinitely small point.

Rather, every telescope takes in a field of view (FOV) of a specific size, such that in all cases there are a finite number of non-overlapping FOVs comprising the total of 41,200 square degrees of sky. Arecibo had a FOV of about 0.002 square degrees at 1.5 GHz. Consequently, it would have taken about 20.6 million beams to observe the whole sky (leaving aside the fact that no ground-based telescope is able to observe the whole sky). The SETI Institute's Allen Telescope Array (ATA), has a much wider beam which could cover the sky in a mere 14,000 pointings, but with much lower sensitivity than Arecibo. Large optical telescopes would require far more pointings to cover the entire sky.

We can retain the nomenclature of $n_s$, but now understand that the "s" denotes "spots on the sky" rather than stars or sources, though it must be appreciated that "spots on the sky" cannot exactly correlate with N because a single ET civilization may occupy more than one FOV or multiple ET civilizations might occupy a single FOV, either in foreground or background. This newly reimagined $n_s$ does, though, have the major advantage of lending itself to an exact calculation for each telescope, thus eliminating the need to vainly speculate, without evidence, as to whether this or that type of star or other target might host ET. It also vastly reduces the burden on SETI scientists. Instead of drawing up endless lists of possible targets, all they need to do is mosaic the sky, FOV by FOV. Since this can be accomplished in 14,000 beams for an ATA class telescope, and if one allots 10 minutes to each pointing, the task can be accomplished in 140,000 minutes, or a mere 97 days, making a full sky SETI survey an ideal use of the ATA. The job would take much longer for an Arecibo-class telescope because of its much smaller beam, about 392 years.

The situation for optical telescopes is the same. Because of the higher frequency of visible light, the most powerful optical telescopes have a smaller FOV than their counterparts in radio astronomy. So, whereas Arecibo would require 20.6 million beams to cover the sky, the 10-meter Keck telescope would require 178 billion pointings to accomplish the same task. Even were the Keck to perform this experiment 12 hours per night, and if 10 minutes were allotted to each pointing, it would take close to 7 million years to get the job done!

There is a solution in the optical if one is willing to sacrifice telescope sensitivity in order to increase FOV. An ET laser should vastly outshine its star (if it is in fact associated with a star) in its specific wavelength. To tease the laser out of the host star's noise one can either count photons per small units of time (e.g., nanoseconds) or spectrographically smear the starlight such that the laser sticks out as a luminous dot. Several wide FOV optical SETI programs are currently planned or underway [9,10,11].

In summary, it may be time to abandon the entire concept of targets, where one is forced to make pointing choices in the absence of any clear reason to prefer one set of coordinates over another, and instead take a theoretically agnostic posture by mosaicking the entire sky, FOV by FOV.

## 3. $f_p$

### 3.1 Planets are Ubiquitous

$f_p$, the fraction of stars with planets, was completely unknown when Drake first devised his equation in 1961. Now it is. For all intents and purposes, all stars can now be deemed to have planets, such that $f_p \sim 1$. This is based upon two main bodies of evidence:

- Images of proto-stars, using optical, radio and infrared telescopes, invariably reveal circumstellar discs of material which presumably indicate a process of planet formation.

- Extrapolations from data collected by the Kepler Space Telescope indicate that planets are ubiquitous.

If planets are ubiquitous, and $f_p \sim 1$, then there is no need to include this factor in the Drake Equation since it does not modify N.

### 3.2 Moons and Dwarf Planets

The case against $f_p$ is enhanced when we realize that a census of planets wrongly neglects moons and dwarf planets. There is no reason that life cannot arise on a moon, and intense speculation has centered on the possibility that life exists on the Jovian moons of Europa and Ganymede, the Saturnian moon of Enceladus, Neptune's moon, Triton, and Uranus' moon, Titania. There is strong evidence that each of these moons harbors a sub-surface liquid water ocean. Dwarf planets, Ceres and Pluto, may also have sub-surface oceans. Add to this, Titan, whose surface flows with hydrocarbons in liquid form, and at a further depth, it too may have a liquid water ocean.

### 3.3 Does Low Metallicity Reduce $f_p$?

Sol is a late generation star. Of course, our star system has high enough metallicity to form rocky planets, since we are sitting on one such. The question arises as to what proportion of stars have enough metallicity to form rocky planets. In the paper most often cited in the SETI literature, Charles Lineweaver estimated that the average age of rocky planets is 6.4 billion years old, and that 74% of all earth-like planets orbiting sun-like stars are older than the Earth [12]. It therefore stands to reason that the first alien civilizations to arise could be billions of years older and more advanced than our own. However, Lineweaver's estimate could be conservative, as it fails to take into account the conditions of the early universe, whose stellar demographics were quite different from the present universe. The early universe contained a vastly larger proportion of massive super-giant stars, which created successive waves of metal producing supernova explosions in a very brief amount of time relative to the age of the universe.

Hence the universe may have been rich enough in metals to harbor rocky planets considerably earlier than Lineweaver estimated. Indeed, recent observations of very ancient white dwarfs have revealed rocky material either in close orbit or on the star's surface, presumably the infallen remains of planets that strayed too close to the white dwarf and were shredded by tidal forces [13,14]. More recent work indicates that metallicity was high enough after approximately the universe's first billion years to form rocky planets, and there is no strong correlation between the number of Earth or larger-than-Earth rocky planets and star metallicity [15]. Moreover, there are few low-metallicity stars at all in at least our neighborhood of the Milky Way.

4.   $n_e$ ➔ $n_{tb}$

$n_e$ represents the average number of Earth-like planets orbiting within the HZ of each star system that bears planets (which we now know to be virtually all stars). Drake took "Earth-like" to mean a rocky planet with liquid water on its surface. Our Sun's HZ is simplistically calculated to lie between 0.7 and 1.3 times Earth's current orbital radius. A comprehensive review of Kepler data by Bryson, et al., concluded that under conservative assumptions for the width of the HZ between 0.37 (+0.48−0.21, errors reflect 68% credible intervals) and 0.60 (+0.90−0.36) planets per star, while under optimistic HZ width assumptions there are between 0.58 (+0.73−0.33) and 0.88 (+1.28−0.51) Earth analogue planets per star. They concluded that the nearest such planet might be as close (on average) as 20 light years away, with about 4 such planets within about 33 light years [16]. This would translate into something on the order of 300 million such planets in the Milky Way.

However, simplistic calculations of the radius of HZ neglect to take into consideration a planet's size, atmospheric pressure and composition, amount of surface water, the presence or absence of volcanism and tectonic crustal movement, among other factors that can make all the difference. By way of a very concrete example, the Earth and the Moon both lie at the same distance from the Sun, yet the former is bursting with life, while the latter is dead. In recognition of this, the HZ is usually constrained further to include that band where surface water would remain liquid in the presence of an Earth-like atmosphere comprised of nitrogen, carbon dioxide and water vapor. However, this is still a great over-simplification. Leaving aside the Moon, Earth itself provides an excellent example of how simple calculations of HZ can mislead. Life formed on Earth at least 3.8 billion years ago at a time when Sol radiated only 70-75% of its present luminosity. That would have placed Earth well beyond Sol's HZ at that time. Its oceans should have been frozen solid. Obviously, the early Earth was habitable. The saving grace is that Earth's atmosphere is known to have been much different then as it is today, being largely devoid of oxygen, but rich in such powerful greenhouse gases as methane, carbon dioxide, sulfur dioxide, carbon monoxide, and water vapor. We know that Mars, which orbited even further outside of the then current HZ, was similarly blessed with a much thicker atmosphere (estimated to have been 100 times thicker than it is presently), and one largely comprised of the greenhouse gas, carbon dioxide. Because of its thick $CO_2$ atmosphere, we know from very much evidence that water flowed on its surface as well.

The concept of the HZ is hopelessly blinkered:

- When Bryson, et al., estimated the number of Earth-like planets in HZs, they did not attempt to include an estimate of the number of moons orbiting larger sub-Neptune or gaseous planets within those same zones.

- It ignores the fact that liquid water can exist below the surface heated by energy sources additional to its star's light. Life as we think we understand it requires a liquid solvent within which nutrients can enter cells, waste products leave, a medium in which internal cellular structures can move about, and which can also catalyze important metabolic functions. Water may be uniquely successful in this role. Hence NASA's oft-repeated mantra, "follow the water" is tantamount to the statement, "we strive to find life." However, the classic HZ ignores the fact that there is a great deal of sub-surface liquid water in the Solar System. Europa has an interior ocean heated mainly by tidal stressing and radioactive decay, and insulated by an estimated 15-25 km. of overlaying ice. That liquid ocean has a depth of 60-150 kilometers, and contains much more liquid water than all the oceans of Earth put together. There may be thermal vents at the bottom of that ocean, much like the thermal vents on Earth around which life has been theorized to have begun. Saturn's moon, Enceladus, also has a tidally caused sub-ice liquid ocean and hydrothermal vents. Evidence gathered by the New Horizons flyby mission indicates that Pluto is quite likely to have a subsurface liquid water ocean as well, perhaps heated by the radioactive decay of elements within Pluto's core.

- The classic HZ ignores atmospheres. It is entirely plausible that life can and does exist in the thick atmospheres of Venus or the gaseous planets, Jupiter, Saturn, Uranus, and Neptune. In each case, there is organic chemistry and strata where temperatures might be temperate enough for life as we know it to exist.

- It ignores life as we do not know it. No one is certain whether life's solvent must be water. Saturn's largest moon, Titan, has a surface temperature of approximately -180° C (-355° F). At these temperatures water freezes to the hardness of granite. Titan, however, features surface lakes that are fed by abundant rainfall and rivers of liquid hydrocarbons, such as ethane and methane. These lakes, thought to contain about 15 times more liquid than Lake Michigan, are filled with a rich soup comprised of multifarious hydrocarbons, including large quantities of vinyl cyanide, known for its tendency to form membranes. All this makes the surface of Titan a candidate for life--just not life as we know it. Although liquid hydrocarbons can act as solvents, what dissolves in them is not identical to what dissolves in water. For example, DNA and RNA work well with water, but they cannot persist within liquid methane. Making Titan an even more intriguing target in the search for life is that it too apparently has a sub-ocean of liquid water.

In fairness, Drake himself and the SETI community at large generally concede that $n_e$ refers to any planet that might harbor life. However if, as is reasonable to surmise, our Solar System's truly possible HZ extends from the clouds of Venus to the oceans of Pluto, and perhaps even beyond, that is, anywhere where organic chemistry is possible, then surely the entire concept of the HZ is ripe for retirement, and $n_e$ can come out of the Drake Equation. Indeed, NASA, other space agencies, and even a private actor, the Breakthrough Foundation, are planning missions to Mars, Europa, Enceladus, and Titan, that is, well outside of the traditional HZ, and doing so explicitly in the quest for life. They are voting with their dollars that the concept of the HZ, as defined by surface water on an Earth-like planet, is obsolete and can be ignored, at least until such time as the entire Solar System has been thoroughly explored for life and determined to be devoid of it.

The $n_e$ term should therefore be exchanged for $n_{tb}$, signifying the average number of total bodies, such as planets, moons, asteroids, comets and Kuiper belt bodies that have a substantial atmosphere and/or liquid water or liquid hydrocarbons, whether those liquids are found on the surface or beneath it. The resulting figure is certain to be vastly larger than the more restrictive $n_e$. One might argue that we can ignore asteroids, Kuiper bodies, and comets because they do not fit into the larger vision of establishing the number of communicating civilizations in the galaxy. After all, even if one grants that life could arise within a 10-mile wide comet, one cannot possibly imagine that technologically intelligent life could arise there. However, rudimentary lifeforms, or their immediate precursors, might begin, *sui generis*, on a comet and then, through panspermia, migrate to a planet or moon that is suitable for technological life.

## 5. $f_l$

$f_l$ represents the fraction of those planets (in our new recasting, the total number of planetary bodies, be they planets, moons, asteroids, Kuiper bodies, or comets), that bear life of any sort. At this time, we have no idea. A big constraint is that we do not today know how life began on Earth. Hypothesized birth places include ponds, tidepools, deep sea hydrothermal vents, and clay interfaces, or it seeded Earth from space. Particularly vexing is that no pathway, even on paper, has been shown to lead organic chemistry inexorably from simple amino acids and sugars toward complex proteins, RNA, and DNA.

Nonetheless, strong arguments can be made to suggest that life is common in the universe; that biology may be a naturally imperative emergent property of chemistry, just as chemistry emerges inexorably, and without invocations of the miraculous, from the laws of physics. Alternative arguments can be made to suggest that life is so rare, that its emergence on Earth was a matter of such wildly improbable freakish chance, that we are effectively alone in the galaxy and perhaps even in the visible universe.

The case for "life is common" begins with the discovery of a large variety of organic molecules in interstellar dust and gas clouds and the denser molecular clouds, such as make up the Orion nebula. Over 100 species of organic molecules have been discovered in interstellar molecular clouds [17, 18], and the prebiotic molecules, $CH_3NCO$ and $HOCH_2CN$, have been detected in the immediate vicinity of protostar Serpens SMM1[19]. Thus, organic chemistry may be baked into star birth. Indeed, it has been observed that the ancient remnants of our star's creation, namely, comets and asteroids, are often rich in organic compounds such as amino acids. For example, a meteorite that landed in Australia in 1974, known as the Murchison meteorite, was found to contain 74 different amino acids, including eight that appear in terrestrial biology. It has long been theorized that much or most of the Earth's water was delivered to it by cometary and asteroid impacts in its early history. At the same time, and by the same means, organic chemistry may have been delivered as well.

Precursors to life were created in a laboratory in the famous Stanley Miller and Harold Urey experiment first performed in 1952. These scientists recreated in test tubes the conditions they presumed to have been prevalent in the atmosphere and oceans of the early Earth. When they filled beakers with ammonia, water, hydrogen and methane, and then applied an electric current, simulating lightning, amino acids accumulated on the bottom of the beaker after only a few days. This experiment has been replicated many times, using different assumptions about early Earth conditions, with similar or better results. Jan Sadownik and colleagues have succeeded in creating fully self-replicating molecules from mixtures of simple chemicals. Crucially, these molecules competed for resources and seemed to evolve in a Darwinian fashion [20]. Although the pathway from very simple organic chemistry to RNA, DNA, and proteins is overwhelmingly daunting, this may not be the case if we think in terms of individual small steps whereby simple replicating—but imperfectly replicating—molecules, compete for raw materials, and where the most efficient molecules differentially self-replicate more often creating new generations of more efficiently replicating molecules, and so on. This is Darwinism applied to pre-biotic molecular chemistry. Where there is replication and variation, evolution can take place. If Darwinian evolution began long before the evolving entities could be considered alive by most definitions, then the spot where life can be defined as such becomes an arbitrarily chosen position on the spectrum of possibilities spanning from inert molecules, to self-replicating molecules, to information bearing molecules, to microbes.

We do not know how life began on Earth, but we do know that it did begin almost at the very first moment at which it became possible, no later than about 700 million years after the formation of the Solar System and the Earth. It is not certain whether life could have started earlier than this as Earth was being pummeled by comets and asteroids in the Late Heavy Bombardment during the period of around 4.1 to 3.8 billion years ago.

Perhaps life did begin earlier than the end of the Late Heavy Bombardment and persisted deep under water or deep underground.  Some zircons of 4.1 billion years of age may show evidence of life.   They certainly provide evidence that water existed on the surface of the planet long before the Late Heavy Bombardment.  What is also certain is that if life did not start earlier, it commenced almost immediately (in geological terms) after the Late Heavy Bombardment had ended.  For example, there are great agglomerations of fossilized stromatolites dating back 3.7 billion years.  William Schopf et al., in examining microfossils found in Australian rocks dated to 3.465 years ago, describe five very distinct species of microbes, including primitive photosynthesizers, methane producers, and methane users.  If life was already that complicated by approximately 3.5 billion years ago, it must have commenced much earlier [20].

There are counterarguments buttressing the case for "life is rare," the strongest of which is anthropomorphic.  It was first made by Brandon Carter [21] and (with modest modifications) goes like this.  *Homo sapiens* exist as living creatures capable of contemplating the universe and conducting SETI experiments.  The Sun has about a 10-billion-year lifetime.  The Sun began at about 70% of its current luminosity and will inexorably continue to brighten in the future.  Although the Sun will not enter its red giant phase, engulfing or nearly engulfing Earth, for another 5 billion years, in a mere 500 million years from the present the Sun will have heated to such an extent that the oceans will boil away and life will cease to exist on its surface.  Since the Sun is about 4.6 billion years old, it seems that humanity, and with it technology, arose only in the final moments of its very possibility.  Had life not started almost as soon as it was possible, it would not have had the time to evolve into technological intelligence.  It may be that life is sufficiently difficult to start that on average it takes much more than a few hundred million years to get going.  Life on Earth may be at the extreme early edge of a bell curve of possible start times.  Since no onset other than an extremely early one can result in the evolution of a technologically intelligent lifeform before solar heating makes it an impossibility, we are effectively alone as a technological species in our galaxy and perhaps even in the visible universe.  Though interesting, this argument has been rebutted [23, 24] and would in any event only apply to stars as luminous as the Sun.  About 95% of stars are smaller and cooler than Sol.  Their lifetimes increase as their size and temperature decrease, such that not a single K or M class star (together making up the vast majority of stars) has in the history of the universe yet expired of its own accord.

Although an exact calculation of $f_l$ may be far off, we may at least have reason to believe that it is either closer to zero or closer to unity within the next few decades.  Were even a single microbe of independent origin from life on Earth discovered anywhere within our Solar System, then it might be reasonably concluded that life in the universe may be ubiquitous, and that $f_l \sim 1$.  If a systematic search of our Solar System finds no other example of *sui generis* life, then the argument for a very low $f_l$ would be strengthened.   There will be quite a few upcoming bites at this apple.

## 5.1 Mars

It has now been determined beyond any doubt, and from many lines of evidence, that water once flowed upon the surface of Mars. For example, the Mars Reconnaissance Orbiter (MRO) has photographed river beds, alluvial plains, lake beds, and gullies; and the Mars rovers, Spirit and Opportunity, have found clays, hydrated sulfates and salts, opals and other minerals that could only form in the presence of water. In examining crater walls, Opportunity found that in places sedimentary rocks had a depth of hundreds of meters, suggesting that the water that deposited the sediments persisted over a very long time.

Although all surface water seems to have disappeared on Mars by about 3 billion years ago, that is still later than the point at which life had begun on Earth under not wholly dissimilar circumstances. There are indications that Mars still harbors underground liquid water. The MRO has observed carved gullies in which dark streaks seem to appear during the Martian spring and summer and disappear in the cold season. The presence of hydrated salts in these gullies lends support to the position that briny water is flowing down the steep gullies.

The unequivocal discovery of Martian fossils would be momentous, but in themselves may not help determine whether Martian life arose in complete isolation from life on Earth but lead to one or the other of two conclusions: Either life is easily initiated; or, once it does begin on one planetary body, it can infect others. Unless Martian life can be determined to be of a completely independent origin, it will not immediately answer the question as to whether life in the universe is rare or ubiquitous. It would likely take a close examination of existing Martian life to determine which is the case. If Martian or any other life form found within the Solar System represents a completely separate biota, then it might be concluded that if life can independently arise two or more times within a single solar system, then it might be assumed to be common or even ubiquitous throughout the galaxy.

## 5.2 Enceladus

Saturn's small moon, Enceladus (diameter = 505 km. or 314 miles), sports a sub-ice ocean that was photographed by the Cassini spacecraft spouting geysers. Because of the moon's very slight gravity, water shoots hundreds of kilometers away from the surface before settling back down on to Enceladus as snow. Over one hundred such geysers were observed by Cassini during its thirteen-year exploration of Saturn and its moons. Those geysers contained organic molecules and trace amounts of methane, which could be indicative of a microbial waste product. The plumes also contain silica nanograins in a configuration that strongly suggests that they derive from hydrothermal vents located on the ocean's rocky bottom. It has long been theorized that life on Earth began at deep sea hydrothermal vents. The Breakthrough Foundation, in possible partnership with NASA, is designing the Enceladus Life Finder mission that would fly through these geysers, testing their water content for microbial life, possibly with an onboard microscope.

## 5.3 Europa

Life as we know it needs water, carbon and other specific elements, plus a useable source of energy.  Europa's surface is entirely covered in water ice to a depth of about 10 - 15 miles.  Below the ice is a liquid ocean containing at least twice the water of all of Earth's oceans combined.   Below that is a rocky core with which that water interfaces.  Photosynthesis is unlikely to work on Europa as the already feeble sunlight that hits Europa would not penetrate the thick ice.  The ocean floor may harbor thermal vents much like those on Earth's seafloor that host and feed a myriad of life forms in the absence of photosynthesis.  Radiation emitted by Jupiter might drive surface chemical reactions that convert carbon molecules into useable food for microbial life below its surface [25].  That organic material might be transported into the water through processes akin to terrestrial tectonic subduction.

NASA plans to launch its Europa Clipper orbiter in 2023.  Its suite of nine instruments will yield many insights into the composition of Europa's ocean, as well as the organic and inorganic materials on its surface and in the water vapor plumes above the moon, through which the orbiter will attempt to fly. The Europa Clipper might be followed by the Europa Lander mission, tentatively scheduled for a 2025 liftoff, which will directly sample the surface of Europa for remnants of life.

## 5.4 Titan

NASA's Dragonfly mission to Saturn's largest moon, Titan, is currently scheduled to launch in 2027.  It features a drone-like instrument that will fly over more than 100 miles of Titan's terrain, sampling its organic chemistry and possibly deducing extant or extinct life from its geological features.

## 5.5  Venus

A new model of Venus suggests that until relatively recently, Venus may have been temperate enough for liquid water to have flowed on its surface.  The era during which surface life might have been possible ended when massive volcanic eruptions covered the planet 1.5 billion years ago with a sea of molten lava and which released enough carbon dioxide into the atmosphere to create the hell-like conditions that we observe today [26].

Phosphine ($PH_3$) has been purportedly discovered in a temperate stratum of the Venusian atmosphere.  There is no known mechanism by which nature produces atmospheric phosphine, except as the waste product of anaerobic microbes [27].   The atmosphere of Venus has an abundance of carbon chemistry as well as water vapor, but also life-as-we-know-it destroying clouds of sulfuric acid.   The purported phosphine signature has been challenged. Further confirmatory observations will be required to settle the matter [28].

### 5.6 Strange Life on Earth

Only a tiny fraction of the microbes on land and especially in the seas have been catalogued. The biomass of all microbes found beneath the earth is estimated to be at least the equal of surface microbes, and yet next to none of these have been identified and studied. We simply cannot say today with certitude whether all life on Earth is descendent from the same ur-life, the so-call "last universal common ancestor" (LUCA). It is possible that life began on Earth several times before or immediately after the Late Heavy Bombardment and that, although the DNA/RNA life we are familiar with dominates the surface of the Earth, other forms retreated to and still hold forth within the depths.

### 5.7 Exo-atmospheres

Kepler and other telescopes have established that as a planet transits its star it slightly, but measurably, dims that star from our point of view. At the point the planet enters and leaves the star's limb, starlight must pass through its atmosphere. The observed spectra will be able to be measured by near future generation telescopes. Oxygen and its byproduct, ozone, would be strong indications of life. Oxygen is highly reactive and therefore if it is detected in more than trace amounts in an atmosphere, a renewable source for it must be postulated. Atmospheric ozone breaks down very quickly. There are currently no known means to create atmospherically significant quantities of oxygen or ozone except biologically. Methane also breaks down quickly, and so it too suggests a renewable source if detected. There are geologic possibilities, but biology may be a better bet, as anaerobic bacteria, such as found in swamps or the guts of cows, are prodigious methane producers. Phosphine may be a very strong indicator of life, especially if the positive detection of it in the Venusian atmosphere is ultimately determined to indicate life. Water vapor, though not an indicator of life itself, may be an important prerequisite gas. In the absence of water vapor, the presence of other possibly bioindicative gases, such as methane, nitrous oxide, and methyl chloride and dimethylsulfide, may be less reliably interpreted as indicating the presence of a biosphere than otherwise.

### 5.8 SETI

In the event that SETI succeeds in making a detection before its sister discipline, astrobiology, the point will have been proven, since technologically intelligent life also counts as life.

### 5.9 Interstellar Panspermia: A Real Wild Card

If an independent genesis of life is discovered within our Solar System then it can be reasonably concluded that life is common throughout the galaxy and the universe. If life is found elsewhere in the Solar System but it is sufficiently similar to life on Earth, then it must be concluded that panspermia is possible. If life started on another planetary body, and then migrated to Earth, this would be fascinating in its own right, but it would not answer the question as to whether life is common in the universe. Life might have resulted from some freak accident peculiar to our Solar System, but that once it did arise, proved resilient enough to be transported from its site of origin to other locations.

The bigger question becomes whether panspermia might function across interstellar distances. If life is capable of migration among star systems, such that we ourselves are of interstellar origins, then life, per force, must be ubiquitous. Microbes and microbial spores that have lain dormant in ocean floor clay sediments for more than 100 million years, and thereby removed from enough oxygen and available carbon to actively metabolize and replicate, have been successfully revived [28]. Alien microbes might have been buried in the subsurface of interstellar asteroids, such as 'Oumuamua.

It is also possible that spontaneously arising life is extraordinarily rare, but that once it does arise it eventually evolves into intelligent life forms such as ourselves. An ur-ET may have resolved to remedy the situation of life's rarity by intentionally infecting the rest of the galaxy by sending small microbe laden probes to nascent solar systems, releasing its cargo upon arrival, and allowing the microbes to evolve on their new home planets or moons however they might [29].

The Drake Equation assumes that life is endogenous, emerging independently on each planet on which it occurs. However, when one also considers the case for panspermia, the frequency with which life arises endogenously on planetary bodies might be overwhelmed by the frequency with which exogenous spores infect them.

## 6. fi

fi denotes the fraction of life bearing planets on which at least one technologically competent intelligent species evolves. The argument has been made by ET-skeptics, such as evolutionary biologists George Simpson [30] and Ernst Mayr [31], that as common as life may or may not be in the universe, the evolutionary pathway that resulted in the emergence of the technologically intelligent species, *homo sapiens*, is so constrained by non-repeatable happenstances that, in effect, we might be alone or nearly alone in at least the galaxy. It has to be conceded that of the billions of species that have ever existed on planet Earth, only one, *homo sapiens,* has achieved technological detectability. The argument goes like this. Had an asteroid not wiped out the dinosaurs 65 million years ago, rat sized mammals of the age would never have emerged from their nocturnal and subterranean existence to live out in the open without fear of being devoured by dinosaurs. Some took to the trees. There they evolved opposable thumbs that aided in grasping tree limbs, with no anticipation that this adaptation would much later prove invaluable in the construction of stone tools and spacecraft.  Sharp color vision was needed to swing from tree to tree, as well as to locate ripe fruit from a distance, in turn requiring larger brains to process the large amount of visual input.  When the forests retreated, leaving our ancestor hominid species to fend for themselves on the savannah floor, they began to eat meat, no doubt starting with small and easy to catch prey.  Meat is more calorically dense than fruit.  Brains require more calories than any other organ as a percentage of body weight.  The modern human brain burns 20% of the body's caloric intake, despite comprising only about 2% of body weight.  Higher caloric intake led to bigger brains, which in turn led to enhanced intelligence, resulting in turn to better hunting skills, leading to the ability to kill larger game, leading to increased caloric intake and on and on in a virtuous feedback loop.  Big game hunting also required social cooperation, which again required greater computing power to manage intricate social cues and relationships, setting up another feedback loop that synergistically reinforced the other one.   Nothing of this evolutionary trajectory was guaranteed.

*Homo sapiens* may very well be a fluke, but the evolution of general intelligence certainly is not.  Problem solving is a vital survival skill that seems to have evolved over and over again, much in the same way as flying has separately evolved in pterosaurs, bats, birds, and insects.  The list of "convergent" adaptations is long and includes eyes, which have separately evolved some 40 different times and sleek torpedo-type body shapes best suited for swimming that separately evolved in fish, squid, dolphins, and penguins among others. Some level of intelligence seems to have evolved in virtually all species.  For example, slime molds and fungal mycelium can be trained to successfully navigate a labyrinth for a nutritional reward.  In fact, just about any microbe displays intelligence in the way it moves toward or away from light, nutrition or danger.  Ants seem to be completely hardwired creatures, but if one stands back and regards an entire colony to be what EO Wilson calls a "super organism," wherein individual ants are analogous to the cells of other animals, the colony acts in a highly intelligent manner, adapting rapidly to all manner of novel threats and opportunities.  Similarly, honey bees, when swarming, will send scouts in various directions to find the best location for a new hive.  When the scouts return to the swarm body, each opines on what she has found through waggle dances, and a collective decision is made.

Dinosaurs persisted on Earth for more than 150 million years, and yet archeologists have yet to uncover a single dinosaur city [32].  Had the asteroid or comet that hit Earth 65 million years ago been delayed by a few hours, it would have missed the Earth entirely and dinosaurs might still walk the Earth today (they are flying the Earth now, in the form of birds).  Although dinosaurs apparently did not develop cities or spacecraft in their first 150 million years, there is no way to be certain that they would not have developed them during the most recent 65 million years.

Exact brain sizes are difficult to determine from fossils, but the small, highly lethal *Dromaeosaurs* appears to have had a brain roughly twice the size of that which might have been predicted by body size alone. It had specialized forearms for seizing prey and may have hunted in packs. If it did hunt in packs, technological intelligence might have evolved from that baseline. Their forearms might have eventually evolved in a manner to better avail toolmaking. Crucially, they appeared only in the last 10 million years of the dinosaur epoch. Similarly, *Strenoychosaurus*, also flourished in the late Cretaceous. This dinosaur was smaller than a human, weighing 60 - 100 pounds, but had a brain that was about 6 times bigger than that of a modern crocodile, as well as large, widely set apart eyes, and primitive hands, possibly used to hunt and grasp small mammals. Certainly, objective observers 65 million years ago would have bet on the *Dromaeosaur* or *Strenoychosaurus* becoming the first to evolve technological intelligence. Few would have bet on the tiny mammals of that time.

Lost also in the standard "but-for-the-asteroid" argument is that mammals did not suddenly burst onto the scene 65 million years ago. The first mammals date to about the same time as the first dinosaurs, and, like them, did not evolve any technological intelligence in their first 150 million years either, but did so only during the very last moment of the 65 million years since that asteroid struck. Finally, when one steps back and takes a very long view of life on Earth one realizes that multicellular animals have only existed in the last approximately 15% of the history of life on Earth. From that perspective, the race from flatworms to scientists was blindingly fast.

It is true that no other species has yet achieved technological abilities, but if humans do disappear, we can easily imagine another species taking its place, perhaps in only a few million years. Various species of birds, cephalopods, dolphins and whales, would be natural contenders but for their lack of manipulative appendages, or the fact that they live in water, or both. Frank Drake, agreeing with Phillip Morrison, suggests that raccoons might be next up. They are omnivores, have very flexible hands, hunt in packs, display highly adaptable behavior, and are keenly intelligent.

7.  $f_c$ ➔ $f_d$

$f_c$ represents the fraction of technologically intelligent aliens that are attempting to communicate with Earth in our present time. It reduces $f_i$ by recognizing that even if a species is technologically competent to do so, it may not because, for example, it is too timid, or it is listening instead of transmitting, or it faces budgetary constraints, or it tried for some long time and has given up in despair, or it is incurious, or its religion forbids it.

The original formulation of $f_c$ is naïve in several respects. It ignores the concept of duty cycle, the amount of time it takes for a transmitting or receiving telescope to make a single circuit through the number of stars or other objects on its target list. If the alien transmitter and our receiver each spend ten minutes every five years directed at the other, then the chances that they will line up in time is miniscule. Unless ET is transmitting isotropically at intrinsically fantastic energy levels, or is constantly transmitting directly to Earth, the $f_c$ contemplated by Drake is actually a vastly lower number than what he and other early SETI thinkers considered.

On the other hand, Drake did not consider that a first detection might not be of an intentionally transmitted signal, but of a byproduct of ET's activity, unrelated to any effort on its part to communicate with Earth. For example, we might detect ET's city lights, or artificial chemicals in its atmosphere, or the atomic particle contrails of its interstellar spacecraft, or spillover from lasers pointed at light sails attached to its interplanetary spacecraft, or the infrared waste heat of large housing, energy farms, or other structures orbiting its star (Dyson spheres), or the unnatural shape (e.g., a triangle) of such structures as they transit its star [33]. When we consider all of these types of "technosignatures," of which intentional efforts to communicate with Earth is but one, the effect is to potentially greatly increase the value of $f_c$ beyond what was contemplated by Drake.

It also fails to consider the possibility that ET is present in our Solar System, surveilling Earth with the intent of communicating once it has decoded our language and culture. In such an eventuality, $f_c$ will for all practical purposes = 1.

Consequently, $f_c$ might be best renamed in a fashion that denotes the probability of detecting aliens with human technology per telescopic pointing or FOV. The replacement factor, $f_d$, departs from $f_c$ in that it is not simply a function of ET's attempts to communicate, but also of our capabilities to detect ET's technosignatures. For example, using existing or near future technology, a fleet of space based optical telescopes acting as an interferometer might be able to detect at least nearby alien city lights. Using this example, $f_d$ becomes a function of the aperture of our telescopes, the size and brightness of ET's cities, and the distance between us. Whereas $f_c$ is a static calculation of the fraction of alien civilizations attempting to communicate with Earth, $f_d$ is a dynamical interface between the respective powers of our two technologies.

## 8. L

Drake intended L to signify the number of years that a technological civilization persists in its attempts to communicate with Earth. However, in more general terms, it is usually taken to mean the average number of years civilizations exist before they self-destruct or meet some other ignoble end.

There is no way to determine L by any other means than to wait until contact is established with ET and ask it what it knows of the matter. Carl Sagan wrote that "all our arguments about the value of $f_l f_i f_c$ pale before our uncertainties in L" [34]. By this he meant that we have actual experience with each of the other factors in the equation. We can know the rate of star birth, we have discovered a plethora of exo-planets, and we know that life exists on at least one planet, and that in us it has become technologically competent and capable of communicating with the stars. However, not knowing our own fate, we can barely speculate about the fate of ET.

That said, speculation persists, and has roughly broken down into two camps-- pessimists and optimists. The earliest SETI theorists wrote during the height of the cold war, in the early 1960's. With nuclear Armageddon on everyone's mind, it was not hard to imagine that L might be as low as 100 years. In those days, before the "agricultural miracles" leading to ample food supplies, famines in one country or another were often in the headlines, while population growth was on a Malthusian trajectory. It was not difficult to assume that civilizations no sooner arise than they self-destruct. Even such early proponents of SETI as Sebastian von Hoerner, John Billingham, and Barney Oliver, felt that as many as 99% of all civilizations would destroy themselves long before they could establish contact with Earth. I. S. Shklovskii, who wrote an early text book on SETI with co-author Carl Sagan [35], eventually reversed himself and renounced SETI altogether on the grounds that civilizations promptly self-destruct before they are able to make contact. Today, a pessimist might look at environmental degradation and climate change and come to a similar conclusion.

Optimists have plenty to point to as well. Steven Pinker's *Enlightenment Now* offers a cornucopia of optimistic thinking [36]. Pinker paints a picture of inexorable worldwide progress across the years, decades, and centuries including large increases in per capita wealth, literacy, life expectancy, attained educational levels; and concomitant decreases in hunger, poverty, violence, child mortality, warfare, numbers of nuclear weapons, and population growth. According to Robin Hanson, there may be a great winnowing filter through which every ET civilization, as well as our own, must pass. The pessimistic view is that the filter lies in our future, and we are doomed. The optimists inherently believe that it is behind us, and that although we may have come through a little the worse for wear, a bright future as a member of a grand galactic club awaits us [37]. It may be that *homo sapiens* are a rare technologically competent species that also just happens to be territorial, tribal, aggressive, and patriarchal. Perhaps the typical ET is none of these, such that a tendency toward self-destruction is extremely rare.

Apart from its inherent unknowability, the L factor is fraught with other problems:

## 8.1 Alien Intentional Transmissions May Not Correlate with L

In practice, the length of time that the civilization persists in transmitting signals may not correspond to the total time that the civilization itself persists. A given civilization might signal for a few thousand years, then stop for a few thousand years while it awaits a response. It may begin a regimen of signaling, only to face budgetary cutbacks as its civilization focuses on other priorities or as its theology changes. A simplistic view of the Drake Equation seems to take for granted that civilizations remain very stable over long periods of time in a way that no Earth civilization has. The generally accepted view that L represents the time that ET is online, that is, actively transmitting signals, is simplistic in another fashion. ET might signal directly to Earth for only a very tiny fraction of the time (or never signal at all) that it is detectable by other means. Therefore, L should be taken more broadly to mean the amount of time an ET civilization is detectable by any means.

### 8.2 L May Not Be Restricted to a Single Species or a Single Civilization

The SETI literature often associates L with the length of time a given communicating civilization persists. However, after the fall of the Babylonian Empire, cows continued to be domesticated, more pottery was thrown, and arrows still flew. In other words, useful technologies almost always outlive specific civilizations, and persist unless and until they are replaced by better technologies. For this reason, the historical ages of humankind are generally denoted by their technologies rather than their civilizations. Thus, we speak of the Stone Age, Bronze Age, Iron Age, Industrial Age, Information Age, rather than the Egyptian Age or Roman Age. Whatever Age supersedes the current, SETI may persist.

In the SETI literature L is sometimes associated not with a civilization but with the average time a species persists. On Earth, mammalian megafauna, such as ourselves, generally face extinction within a timeframe of roughly a million years. However, this may not have strict bearing upon L because not all species go extinct so much as evolve into something else. *Homo erectus* may not walk the Earth today, but presumably evolved into *homo sapiens*. Its tool making skills did not perish with the species, but were passed along and improved upon. Whatever *homo sapiens* might evolve into, the new species (be it biological or artificial) will probably be capable of conducting SETI if it chooses.

### 8.3 Drake's L Ignores the Effect of Possible Interstellar Migration

A single successful and long-lived civilization might spawn many more. Interstellar colonization was never considered in the original Drake Equation, though others have since modified the equation with the addition of a colonization factor [38, 39]. Nor did Drake consider the case for artificial beings. The galaxy may be densely populated with intelligent machines, which may be capable of self-replication (so-called von Neumann replicators) [40]. Requiring neither food, nor water, nor sleep, nor any other life support, apart from shielding from radiation and micrometeorites, plus an energy source, they could course the galaxy indefinitely, residing in orbit around any star, or anywhere in between.

### 8.4 Drake's L Ignores the Possible Role of Local ET Probes

In 1960, a year before Drake wrote his equation, Bracewell introduced the idea that ET might send probes throughout the galaxy [39]. The idea has been fleshed out by others [41,42,43,44,45]. I have postulated that the local probes might be in contact with communication nodes that may be located throughout the galaxy [5,6,7]. The probes and nodes hypothesis completely obviates the need for an L factor, since once probes and nodes are launched they might act autonomously, regardless of whether their progenitor civilization persisted.

### 8.5 The First ET Detection May Be of an Extinct Alien Species

It is entirely possible that we might first detect some other alien artifact other than an active probe or node within our Solar System, possibly one that is information rich. Given that its progenitor civilization may be long extinct, it is unclear whether it should count in the determination of N, which purportedly only counts extant civilizations.

### 8.6 N = L

The ultimate futility of the Drake Equation is summed up best by Frank Drake's own vanity license plate, NEqlsL (N = L), which is a very reduced Drake Equation indeed. Drake estimated that the large number of suitable planets would be cancelled out by the aggregate of the small fractions that one might speculatively assign to $f_l$, $f_i$, and $f_c$.

If Drake is correct that N = L, since L is utterly unknowable by any other means than asking ET, and we cannot make that inquiry unless and until aliens deign to communicate, then the whole Drake Equation may be rendered nugatory.

## 9. A RADICAL RESTATEMENT AND REDUCTION OF THE DRAKE EQUATION

This then is the reshaped Drake Equation, given here as a strawman. For reasons stated below, it cannot actually be computed:

$$N = n_s\, f_p\, n_{tb}\, f_l\, f_i\, f_d\, L \tag{3}$$

Where,

$n_s$ = number of spots on the sky, or FOVs.

$f_p$ = fraction of stars with planets.

$n_{tb}$ = average total number of bodies within each solar system that could engender life.

$f_l$ = the fraction of those that actually do give birth to life.

$f_i$ = the fraction of solar systems with life that evolves technological intelligence.

$f_d$ = the fraction of technological life that is detectable by any means.

L = the duration of detectability.

Because $f_p$ ~ 1, it can come out of the equation.

Because $n_{tb}$ is a very large number, more than a trillion in the case of our solar system if one considers every comet, asteroid, icy moon and dwarf planet, and each terrestrial hydrothermal vent, tide pool, puddle or pond as its own petri dish, each an independent experiment where life might, by chance, have emerged just once from prebiotic chemistry. Consequently, either $f_l$ (the fraction of star systems that are biotic) is also ~ 1, if life is anything more than virtually miraculous, or it is almost zero in the event that life is so incredibly difficult to cook up that given even more than a trillion possibilities per solar system it almost always fails to emerge. Once any factor is assigned a zero, then N collapses to zero as well.

Therefore, we will set $n_{tb} f_l$, (considered together) as ~ 1, otherwise SETI would be a pointless endeavor. Taken together, being ~1, they do not assist in reducing N and therefore both factors can come out of the equation.

No matter how far the intelligence of squid, orcas, or ravens might evolve in Earth's future, they are unlikely to ever become detectable by virtue of any product of their intelligence, namely, technology. A technological civilization that cannot be detected because its technology is too weak (e.g., its city lights are not bright enough) relative to its distance from Earth and the power of our detectors, from the point of view of SETI, is no different than a flock of ravens and just as irrelevant in determining N. If the product of its intelligence is detectable by any means, it is already counted under $f_d$. Therefore, $f_i$ as an independent variable can come out of the Drake Equation.

Drake's L can also come out of the equation. Because of the reasons listed in Section 8 above, a detection of ET may come from outside the bounds of ET's actual temporal existence. There may be a correlation between N and L, but it may be low. Furthermore, L may be irrelevant if what is of interest to a SETI scientist is the probability of detecting ET with each telescopic pointing or beam. In effect, $f_d$ is a stand in for L, or redundant with it.

This leaves $f_d$. Since it now stands alone, it can be further defined as the probability of detecting aliens with each telescopic pointing, beam or FOV. However,

$$N \neq f_d \qquad (4)$$

in that the probability of detecting ET within each FOV only very loosely correlates with the actual number of extant alien civilizations in the Milky Way.

N itself cannot be determined by any means we can currently devise other than by entering into a dialog with ET and asking what it might know of the matter.

## 10. CONCLUSIONS

The multifarious shortcomings of the Drake Equation have been addressed by others [46,47], while its overall failure of the Drake Equation has been exposed by the many prior attempts to solve it, which have resulted in solutions that range over fully 8 orders of magnitude [48]. Drake reduced his equation to N = L, and here it is reduced further to a single standalone factor, $f_d$. Were the new factor helpful in solving for N, astronomers and the public might pay it great homage, perhaps by calling it the Gertz Factor. Unfortunately for its would-be fame, $f_d$, (like Drake's L) cannot be determined by any means other than observation. It is precisely because neither the Drake Equation nor the new Gertz Factor can be solved in the absence of observations that SETI is so important. Moreover, even a vain attempt to solve for Drake's N or determine $f_d$, in the absence of observations, makes no practical difference to SETI scientists. They are still required to plow through targets one by one, or, as suggested in this paper, mosaic the entire sky FOV by FOV. The Drake Equation provided early SETI thinkers with an extraordinarily useful heuristic to begin the serious consideration of SETI. However, this aging edifice now for the most part tends to hinder fresh thinking.

## 11. ACKNOWLEDGEMENTS

The author gratefully acknowledges the kind assistance of Geoff Marcy in the organization of the ideas expressed in this paper as well as Andrew Siemion.